\documentclass[twocolumn,aps,pre,showpacs,preprintnumbers,amsmath,amssymb, superscriptaddress]{revtex4}

\usepackage{graphicx}
\usepackage{dcolumn}
\usepackage{bm}
\usepackage{longtable}

\begin{document}


\title{Percolation through Voids around Toroidal Inclusions} 


\author{A. Ballow}
\author{P. Linton}
\author{D. J. Priour, Jr}
\affiliation{Department of Physics \& Astronomy, Youngstown State University, Youngstown, OH 44555, USA}


\date{\today}

\begin{abstract}
In the case of media comprised of impermeable particles, fluid flows through voids around impenetrable grains.
For sufficiently low concentrations of the latter, spaces around grains join to allow transport on macroscopic scales,
whereas greater impenetrable inclusion densities disrupt void networks and block macroscopic fluid flow.  A critical
grain concentration $\rho_{c}$ marks the percolation transition or phase boundary separating these two regimes.  With a 
dynamical infiltration technique in which virtual tracer particles explore void spaces, we calculate 
critical grain concentrations for randomly placed interpenetrating impermeable toroidal inclusions; the latter consist of surfaces 
of revolution with circular and square cross sections.  In this manner, we study for the 
first time continuum percolation transitions involving non-convex grains.  As the radius of 
revolution increases relative to the length scale of the torus cross section, the tori develop a central hole, a 
topological transition accompanied by a cusp in the critical porosity for percolation.    
With a further increase in the radius of revolution, as constituent grains become more ring-like in appearance, 
we find that the critical porosity converges to that of high aspect ratio cylindrical counterparts only for randomly oriented 
grains. 
\end{abstract}

\maketitle
\section{Introduction}

Fluid flow and charge transport in porous media is of practical and fundamental importance in a variety of settings.
In the case of materials comprised of impermeable grains, fluid flow is through irregularly shaped void spaces surrounding the 
impenetrable particles rather than through well delineated channels.  The concentration of the constituent grains determines 
the degree of fluid flow or charge transport in the medium.  With increasing volume density $\rho$ ($N/V$) of impermeable barrier 
particles, void regions among grains are more likely to be smaller truncated volumes with no access to the rest 
of the system.  Bulk level transport ultimately 
ceases beyond a critical concentration $\rho_{c}$.  The shift from system spanning void networks admitting 
fluid flow on a macroscopic scale to isolated volumes whose finite size precludes bulk permeability is a percolation transition,
a genuine second order phase transition with all of the concomitant hallmarks and singular behavior~\cite{stauffer}.
The percolation threshold is customarily specified with 
$\eta_{c} = \rho_{c} v_{\mathrm{B}}$ and $\phi_{c} = e^{-\eta_{c}}$ ($v_{\mathrm{B}}$ being the volume of the 
interpenetrating grains), with $\phi_{c}$ being the critical or void volume fraction. 

Discrete percolation phenomena as well as continuum percolation involving overlapping particles are amenable to a variety of 
techniques such as the Hoshen-Kopelman algorithm~\cite{hoshen} for identifying connected clusters and thereby determining if a system percolates.
However, while the flow of fluid through spaces between inclusions offers a natural way to describe percolation phenomena for 
porous materials made up of impenetrable grains, the geometry of networks of connected voids is in general difficult to anticipate
a priori.  

Void percolation transitions have previously 
been studied in the context of interpenetrating particles with a variety of geometries.  
Due to the high degree of symmetry, determining if assemblies of randomly placed spheres 
contain system spanning contigous void volumes is amenable to 
Voronoi tesselation~\cite{elam,marck,rintoul,klatt}.  More generally, 
extrapolation to the continuum limit of discretized systems with ever finer mesh has been used in the case of 
randomly placed spheres~\cite{martys,maier}, ellipsoids~\cite{Yi1,Yi2}, and aligned cubes~\cite{koza}.
For a geometrically exact approach applicable to a broader range of grain shapes, we 
use virtual tracer particles to dynamically infiltrate void spaces, where interactions with grains involve specular 
reflections off of the surfaces of impermeable inclusions.  Dynamical simulations in this fashion 
have been reported previously in the context of randomly placed spheres~\cite{hofling,spanner,hofling2,kammerer,spanner2,djpriour}
and for the case of convex solids such as the platonic solids, cylinders, cones, and ellipsoids~\cite{djpriour2}.
For a computationally efficient large-scale treatment, we report on simulations in which virtual tracer particles begin 
their journeys at the center of a large cubic assembly of grains, with linear trajectories interrupted by collisions with and 
specular reflection from the surfaces of impenetrable inclusions 
(i.e. as in the Lorentz gas model~\cite{hofling,lorentz,bruin,beijeren,bauer}).

\begin{figure}
\includegraphics[width=.45\textwidth]{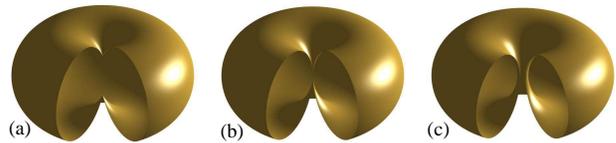}
\caption{\label{fig:Fig1} (Color online) Cutaway view of tori of circular cross section with $\tilde{r}_{1} = 0.47$ in panel (a),
$\tilde{r}_{1} = 0.5$ in panel (b), and $\tilde{r}_{1} = 0.53$ in panel (c)}
\end{figure}

\begin{figure}
\includegraphics[width=.45\textwidth]{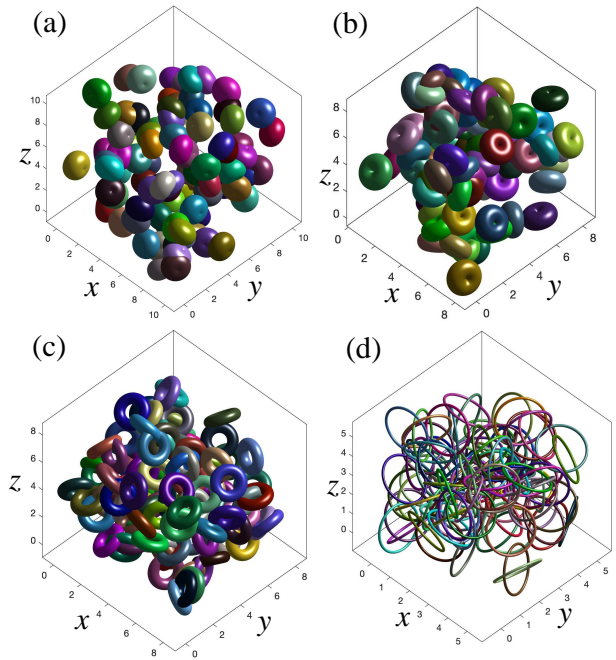}
\caption{\label{fig:Fig2} (Color online) Randomly oriented tori for $\tilde{r}_{1} = 0.30$ in panel (a), 
$\tilde{r}_{1} = 0.50$ in panel (b), $\tilde{r}_{1} = 0.70$ in panel (c), and $\tilde{r}_{1} = 0.95$ in panel (d)}
\end{figure}

\begin{figure}
\includegraphics[width=.45\textwidth]{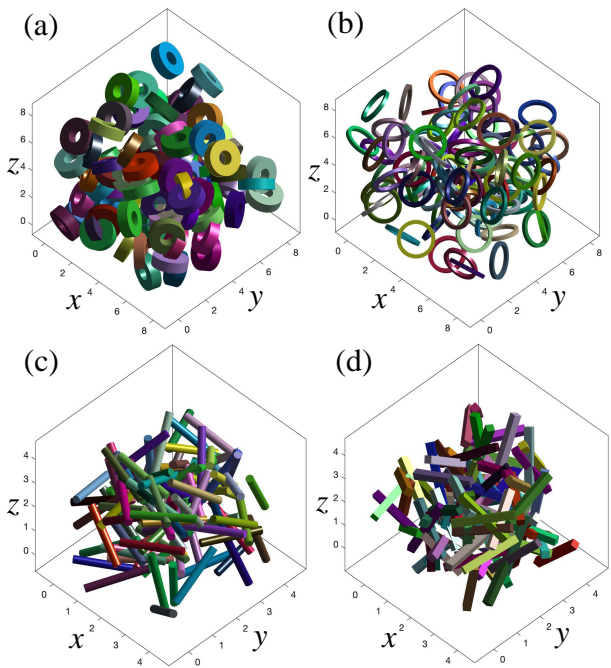}
\caption{\label{fig:Fig3} (Color online) Randomly oriented tori of square cross section for $r_{1}^{*} = 0.70$ in panel (a),
$r_{1}^{*} = 0.90$ in panel (b), cylinders for $r_{\mathrm{A}} = 8$ in panel (c)
and square prisms for $r_{\mathrm{A}} = 8$ in panel (d).}
\end{figure}

Common to previous studies of void percolation phenomena is the convexity of constituent grains.  For the first time, to our
knowledge, we calculate percolation thresholds and study critical behavior for situations in which the impermeable particles
are non-convex.  In particular, we consider percolation phenomena involving torus-shaped grains bounded by surfaces
of revolution about an axis of symmetry (i.e. tori with square and circular cross section).  Moreover, in this work 
we examine porous media made up of randomly oriented inclusions as well as assemblies of tori with their axes of 
symmetry aligned.
In addition to the novelty inherent in their manifestly non-convex geometry, 
the study of toroidal grains offers the opportunity to examine how void percolation is influenced by 
the change in topology
as the genus increases from zero to one.   This transition is marked by the appearance of a central 
hole as solid cylinders become annular cylinders for tori with square cross sections while a symmetrically placed 
pair of cone shaped indentations on a spheroidal surface above and below the equatorial plane merge and yield a ``doughnut'' shape for tori with circular 
cross sections as illustrated in Fig.~\ref{fig:Fig1}.  In the case of the latter, singular behavior in the 
form of a cusp in the critical porosity fraction
accompanies the appearance of a central hole.
Examples of porous media considered in this work appear in Fig.~\ref{fig:Fig2} for toroidal grains of circular cross section and 
in panels (a) and (b) of Fig.~\ref{fig:Fig3} for tori of square cross section.  Linear counterparts for the former and latter 
include cylinders and square prisms (e.g. as shown in panels (c) and (d) of Fig.~\ref{fig:Fig3}).  Critical porosities 
for cylinders have been reported on previously~\cite{djpriour2}, and we calculate percolation thresholds for square prisms 
in this work. 

We label the regime in which the radius of revolution is very large 
relative to the cross section diameter or 
width as the high aspect ratio limit since cutting and distorting the ring-like toroidal grains into a 
linear shape yields very narrow square prisms or cylinders in this regime. 
For the sake of a direct comparison, the aspect ratio $r_{\mathrm{A}}$ for cylinders and square prisms is the 
ratio of the length to diameter or square side width.  On the other hand, in the case of tori, $r_{\mathrm{A}} = \pi r_{1}/r_{2}$,
or the ratio of the circumference of the circle defined by the centers of the shapes of revolution to the cross section diameter or 
width, as appropriate.
We calculate percolation thresholds 
for aligned and randomly oriented high aspect ratio toroidal grains as well as for the linear counterparts,   
finding a convergence of critical porosities to a finite value.  That small segments of high aspect ratio
tori resemble their linear counterparts suggest at least the possibility of a convergence of their critical porosities in this regime.  
We find convergence to a common critical 
porosity only in the case of randomly oriented grains, while $\phi_{c}$ for  tori with their axes of symmetry aligned is 
significantly higher than that of aligned linear counterparts in the high aspect ratio limit.  In the case of the latter, the 
discrepancy is likely due to at least in part to the fact that in spite of the alignment of the axes of symmetry, the random orientation 
of small quasi-linear segments of neighboring
ring-like tori in the torus equatorial plane precludes the presence of long channels among aligned square prisms or cylinders
of circular cross section that would otherwise facilitate fluid flow on macroscopic scales.  

\section{Methods and Techniques}

\subsection{Overview}

Single parameter finite size scaling analysis has historically been an important tool in the study of critical behavior 
in percolating systems and continuous phase transitions more broadly.  In this framework, finite size effects facilitate 
rather than impede extrapolation to the thermodynamic limit, and central to the analysis is the scaling form 
$F(L,\rho) = L^{\mathrm{A}/\nu} g[L^{1/\nu} (\rho - \rho_{c})]$ where $F$ is a thermodynamic observable, 
$g$ is a universal scaling function, $L$ is the system size,
while $A$ and $\nu$ are critical exponents associated with $F$ and the correlation length $\xi$ respectively.
This scaling analysis on a spatial domain draws on knowledge of void networks not often readily available, and for this 
reason we operate instead on the temporal domain; we calculate time dependent observables such as the Root Mean Square (RMS)
displacement $\delta_{\mathrm{rms}}$ from a randomly selected interstitial point at the center of a large cube shaped 
simulation volume.  We consider a scaling form 
$\delta_{\mathrm{rms}}(\rho,t) = t^{k} r[t^{x} (\rho - \rho_{c})$]~\cite{stauffer,avraham} where $k = 1/d^{'}_{\mathrm{w}}$ and 
$x = 1/d_{\mathrm{w}}$ are anomalous diffusion exponents for unrestricted motion averaged over all void 
networks for $k$ and  diffusion only through system spanning void clusters for $x$. Here, $d^{'}_{\mathrm{w}}$ 
and $d_{\mathrm{w}}$ are fractal dimensions of the former and latter random walks~\cite{stauffer,avraham}; $r(y)$, where 
$y \equiv t^{x} (\rho - \rho_{c})$, is a scaling function.  
Universal scaling arguments yield $k = (\nu - \beta/2)/(2 \nu + \mu - \beta)$
and $x = 1/(2 \nu + \mu - \beta)$ where $\nu = 0.8764(12)$~\cite{wang}, $\beta = 0.41810(57)$~\cite{wang}, and $\mu$ are universal critical 
exponents for percolation phase transitions.  Although the dynamical exponents $k$ and $x$ may deviate 
from their discrete lattice counterparts $k_{\mathrm{lat}} = 0.2001(7)$ and $x_{\mathrm{lat}} = 0.2998(4)$
(obtained using $\mu_{\mathrm{lat}} = 2.0009(10)$~\cite{kozlov}), universality theoretical arguments posit that 
$k/x = \nu - \beta/2 = 0.667(1)$.

\subsection{Dynamical Infiltration}

To implement the diffusive exploration of void networks, we use a geometrically exact dynamical infiltration technique   
in which virtual tracer particles follow linear paths and interact with impenetrable inclusions via specular reflections. 
This approach, validated in the context of porous media comprised of a variety of convex  
grain shapes also has the advantage of requiring only information local to the path of the tracers, allowing 
access to larger system sizes while permitting a computationally efficient approach. 
The principal observable we calculate is $\delta_{\mathrm{rms}}$, the RMS displacement from the start of the tracer trajectory.  
The latter encompasses on average at least $10^{7}$ scattering events from impermeable grains, and we consider 50,000 
disorder realizations for each inclusion concentration we sample.
To optimize computational efficiency, 
both in terms of the speed of the calculations and memory usage, we subdivide the simulation volume into cube shaped voxels 
(similar to Verlet cells~\cite{bruin}).  The virtual tracers propagate along their linear trajectories  
given by $\vec{x} = \vec{x}_{0} + \hat{v}t$ with $\vec{x}_{0}$ being the starting point and $|\hat{v}| = 1$ for the (unit) 
velocity magnitude until either reaching the 
nearest voxel wall or collide with a grain, whichever is closest.  In the event of the latter, specular reflection is
implemented and the virtual tracer trajectory resumes with a new velocity 
$\hat{v}_{\mathrm{new}} = \hat{v}_{\mathrm{old}} - 2 (\hat{v}_{\mathrm{old}} \cdot \hat{n}_{\mathrm{loc}}) \hat{n}_{\mathrm{loc}}$ 
where $\hat{n}_{\mathrm{loc}}$ is the local 
(unit) normal direction at the intersection point and $\hat{v}_{\mathrm{old}}$ was the previous velocity. 

An overall rescaling of the simulation volume and its contents has no impact on the status of the system 
with respect to percolation since $\eta = \rho v_{\mathrm{A}}$ remains constant. Nonetheless in practical terms, 
a change in the relative size of grains and the voxels is consequential due to the effect on 
the computational burden of propagation of tracers through the void network.  
If the impermeable inclusions are too small relative to the voxels, the large number of grains contained or shared
by cubic cells erodes
the advantage of dividing the system into smaller sub-volumes.  
In the other extreme, with voxels dwarfed by very large grains,
the benefit of having very few grains overlap a cube shaped cell 
is offset by the small distances traversed by virtual tracer particles as 
they travel from one cubic cell to the next.  
Rescaling the grains while the voxel dimensions (with unit edge lengths) remain fixed,  
we find computational efficiency to be optimized when on average on the order of a 
dozen inclusions overlap with a voxel.

It is convenient to consider spheres circumscribed about inclusions.  
In checking for interactions with impermeable inclusions, many candidates are eliminated 
with a comparatively small computational investment
if spheres circumscribing the grain are not penetrated by tracer trajectories, with the 
task of finding intersections with the geometrically more intricate inclusions themselves 
reserved for the small share of candidates not ruled out in this manner. In addition, the sphere radius $r_{\mathrm{s}}$ is comparable to
the grain length scale thus serves as a dilation factor, scaled up from unity as needed to 
keep the number of grains in contact with a voxel on the order of a dozen.  

\begin{figure}
\includegraphics[width=.45\textwidth]{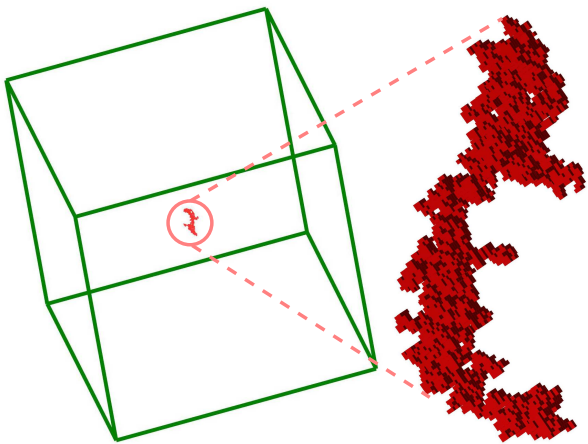}
\caption{\label{fig:Fig4} (Color online) Voxels visited by tracer particles shown in red for $10^{7}$ 
interactions of virtual tracer particles with grains for tori of circular cross section with $\tilde{r}_{1} = 0.70$.
Small red cubes are voxels visited by tracer particles, and the simulation volume is indicated by the large green cube.}
\end{figure}

We operate in the vicinity of the percolation transition where system spanning void networks encompass only a small 
fraction of the total interstitial volume in the porous medium, of which an even smaller sub-volume is 
explored by virtual tracers.  A typical case with only voxels visited by tracer particles highlighted (small red cubes) is shown 
in Fig.~\ref{fig:Fig4} where the large green cube indicates the simulation volume as a whole. 
The fact that tracer particles only infiltrate a small portion of the simulation 
volume offers a significant computational advantage, which we exploit by only requiring that voxels visited by the tracers and neighboring 
cells (i.e. voxels which in principle could contain a grain in contact the cell occupied by the tracer)
be populated with randomly placed impermeable inclusions.  

The manner in which voxels are populated differs depending on whether the cell is invaded by a virtual tracer or is neighbor to a voxel 
entered by a tracer particle. Voxels are deemed ``blank'' until impermeable grains are introduced by rigorously sampling 
Poissonian statistics which can be achieved with a lookup table in conjunction with stochastic input.  
If a virtual tracer enters a blank cell, then it and each of the neighboring voxels are populated with 
inclusions if they are themselves blank. Subsequently, all grains in contact with the cell occupied by the tracer particle are 
identified (along with those in the cell itself) in the event that the 
tracer returns to the same cell as frequently occurs.  On the other hand, 
if a virtual tracer enters a voxel previously populated as a neighboring cell, one need only 
populate blank cells which neighbor it with 
randomly placed grains and identify inclusions in contact with the voxel.  Operating in this way
with a suitable rescaling of the grain dimensions significantly 
improves computational efficiency, particularly in the high aspect ratio regime, while also reducing memory usage to the degree that 
we consider in this work cubic simulation volumes 1000 unit lengths on a side and containing $10^{9}$ voxels 
(each in contact with on the order of a dozen inclusions) in total.  

In this work, we calculate percolation thresholds and study critical behavior for tori with circular and square cross 
sections.  For the latter, we also report results for the linear counterpart, square prisms; critical porosities for 
cylinders with circular cross sections have been previously calculated~\cite{djpriour2}. Finding intersections of tracer trajectories with 
candidate impermeable grains and identifying neighboring grains in contact with a cube-shaped voxel draws on details of the inclusion geometry.
Virtual tracers diffusing through void networks interact with the nearest impermeable inclusion, 
tantamount to finding the minimum travel time for the trajectory $\vec{x} = \vec{x}_{0} + \hat{v} t$.  

In the case of square prisms,
one need only for intersections with each of the six planar facets, subject to the constraint that the intersection point is 
interior with respect to the five remaining planes.  Finding intersections of tracer trajectories
with non-convex grains such as the toroidal inclusions we consider in this work is geometrically more subtle.
The randomly placed toroidal inclusions we consider in this study are uniquely characterized by the orientation of the axis
of symmetry, the location of their geometric center, and the length scales $r_{1}$ and $r_{2}$.  For tori with square
and circular cross sections, $r_{2}$ is the cross section radius for the latter and half the edge length of the square cross section
for the former while $r_{1}$ is the radius of revolution in both cases.
It is convenient to specify tori of circular cross section  
in terms of the dimensionless variable $\tilde{r}_{1} = r_{1}/(r_{1} + r_{2})$, where $\tilde{r}_{1} = 0$ corresponds to 
spherical grains, while $\tilde{r}_{1} = 1$ is the high aspect ratio ring-like limit for inclusions, and $\tilde{r}_{1} = 0.50$ marks the 
topological transition where the central hole appears with increasing $\tilde{r}_{1}$.  
Similarly, in the case of tori of square cross section we use $r_{1}^{*} = r_{1}/r_{\mathrm{s}}$, where $r_{\mathrm{s}}$ is the radius of 
the circumscribed sphere, or $r_{1}^{*} = 1/\sqrt{1 + (r_{2}/r_{1}) + (r_{2}/r_{1})^{2} }$.  As in the case of tori of circular cross 
section, the extreme ring-like limit is attained as one approaches $r_{1}^{*} = 1$, while a cylindrical central channel emerges for $r_{1}^{*} = 1/\sqrt{5}$.

As non-convex objects, toroidal inclusions allow for the possibility of multiple specular reflections from the same grain, an element of the 
tracer dynamics which must be taken into consideration both for tori of circular and square cross sections.  The former are annular 
cylinders with an upper and lower planar surface and inner and outer curved surfaces.  Only tracers scattered from the inner boundary 
may interact in immediate succession with the same inclusion, bouncing off the wall of the central hole until emerging from the cylindrical channel.
On the other hand, tori with circular cross sections are bounded by a single smooth surface and the tracer path may 
cross the torus as many as four times or may avoid contact with it altogether.  As discussed in the Appendix, this circumstance is 
unsurprising given that solving for $t$ involves solving a quartic equation, which one may do analytically. 

Essential to the very modest scaling of the computational burden (given an optimal choice 
of the dilation factor $r_{\mathrm{s}}$) with the grain aspect ratio is finding which inclusions in 
neighboring voxels share the volume of the cell occupied by the tracer.  In the case of square prisms, one checks if any of the eight 
prism vertices are interior to the voxel and vice versa for the cell vertices and the prism interior, which would indicate 
overlap among the voxel and the inclusion.  Similarly, edge segments of either polyhedron penetrating the interior of the other also 
indicate contact among the cell and the prism shaped grain.

For inclusions with toroidal geometry, a straightforward and tractable way to check for overlap with a voxel is to circumscribe
the cubic cell with a sphere and check for overlap among the sphere and the toroidal inclusions.  For this 
purpose, one operates in a coordinate system with the torus geometric center at the origin and the $z$ axis aligned with 
the grain's axis of symmetry.  One then chooses the the $x$ axis such that the $xz$ plane bisects the sphere 
circumscribed about the voxel.  In this manner, determining if the sphere and the torus overlap is reduced to
the task of checking to see if a circle overlaps two squares or two circles for tori with square and circular 
cross sections, respectively. 

To calculate the percolation thresholds and study critical behavior, we conduct dynamical infiltration simulations for 
grain densities $\rho$ in the vicinity of $\rho_{c}$.  Dwell times for virtual tracers are such that tracers interact with 
impermeable inclusions at least on the order of $10^{7}$ times.  Critical concentration results are in accord with an 
independent set of simulations with an average of $10^{6}$ collisions.

\subsection{Single Parameter Temporal Scaling Analysis}

\begin{figure}
\includegraphics[width=.45\textwidth]{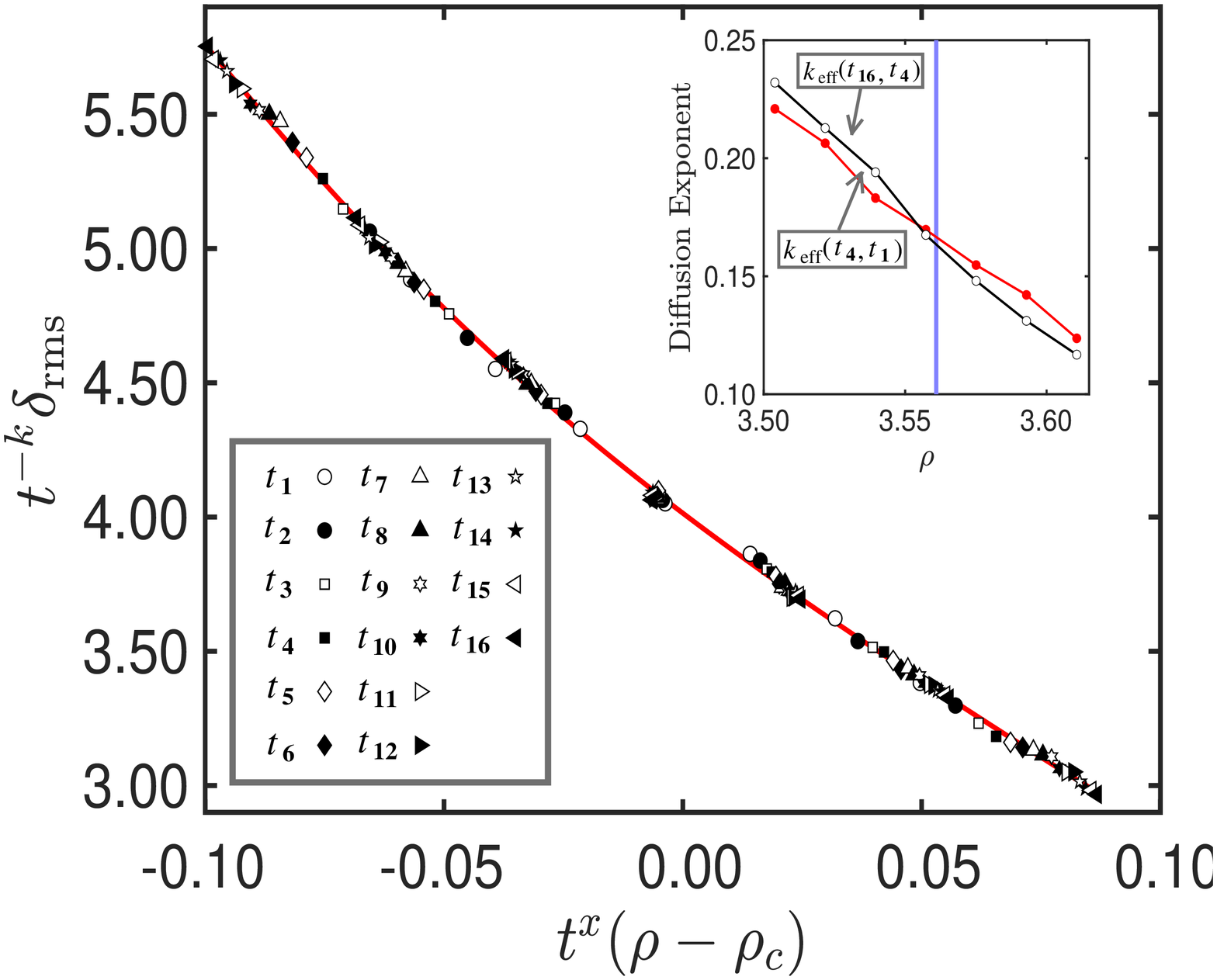}
\caption{\label{fig:Fig5} Sample data collapse for randomly oriented tori of circular cross section for $\tilde{r}_{1} = 0.75$.
The red line is an analytical curve while symbols represent Monte Carlo data. The inset shows 
a crossing of diffusion exponents with the vertical blue line indicating $\rho_{c}$ obtained from a quantitative data collapse.}
\end{figure}

We use a scaling analysis with respect to time to calculate the critical indices
$\rho_{c}$, $x$, and $k$. Much as is done in the context of finite size scaling analysis, we 
exploit the data collapse phenomenon in which Monte Carlo data in principle fall on a single curve when
plotted with respect to $y = t^{x} (\rho - \rho_{c} )$ [i.e. the argument of the scaling function $r(y)$] 
for $\rho$ in the vicinity of $\rho_{c}$.
With a technique described elsewhere~\cite{djpriour}, we perform the data collapse systematically and 
objectively by considering a scaling function $r(y) = \sum_{j = 1}^{n} A_{j} y^{j}$, and optimizing with 
respect to the $A_{j}$ coefficients via linear least squares fitting for a given choice of 
$\rho_{c}$, $x$, and $k$.  We then stochastically vary candidate values for the 
latter three critical indices until the fit of the 
analytical scaling to the Monte Carlo data is optimized in a least squares sense.
The main graph in Fig.~\ref{fig:Fig5} shows an example of a data collapse 
generated in this manner.

As an alternative approach, one may also define effective diffusion exponents with 
$k_{\mathrm{eff}}(t_{j},t_{i}) \equiv \ln [\delta_{\mathrm{rms}}(t_{j})/\delta_{\mathrm{rms}}(t_{i})]/\ln(t_{j}/t_{i})$, or 
the secant line for the Log-Log curve of $\delta_{\mathrm{rms}}$ over the time interval $[t_{i},t_{j}]$.
Percolation thresholds indicated by the crossings are in accord with data collapse results up to the Monte Carlo
statistical error.  The inset of Fig.~\ref{fig:Fig5} displays a crossing of effective diffusion exponents with the vertical 
blue line representing the critical concentration gleaned from a quantitative data collapse.

\section{Results and Discussion}

\subsection{Percolation Thresholds and Critical Exponents}

\begin{figure}
\includegraphics[width=.45\textwidth]{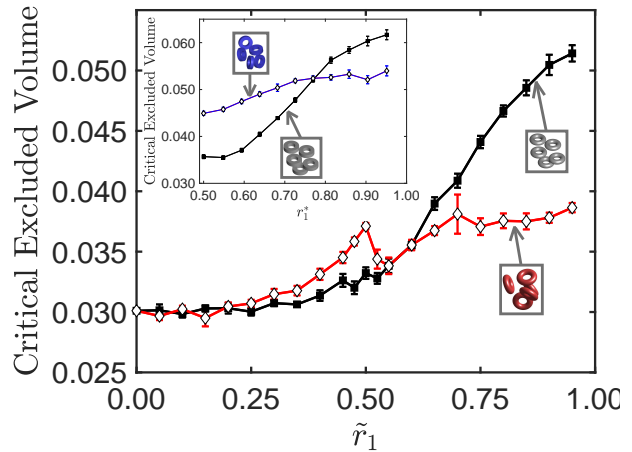}
\caption{\label{fig:Fig6} (Color online) Critical Excluded volumes for aligned (filled square symbols) and randomly oriented (open diamonds) tori of 
circular cross section in the main graph and tori of square cross section in the graph inset.}
\end{figure}

In Fig.~\ref{fig:Fig6}, $\phi_{c}$ results are shown with respect to $\tilde{r}_{1}$ for aligned and randomly oriented tori.  The main graph and inset
show results in the case of tori with circular and square cross sections respectively, and critical indices are
recorded in Tables~\ref{tab:Tab1},~\ref{tab:Tab2},~\ref{tab:Tab3}, and~\ref{tab:Tab4}.  Filled squares correspond to aligned toroidal grains,
while open diamonds represent the randomly oriented counterparts.  In the case of tori with circular cross sections, a prominent cusp 
in the $\phi_{c}$ curve is evident for randomly oriented inclusions for $\tilde{r}_{1} = 0.5$, coinciding with the topological transition marked
by the emergence of a central hole with increasing $\tilde{r}_{1}$.  For $\tilde{r}_{1} > 0.5$, the critical porosity drops sharply before gradually 
rising, appearing to saturate at $\phi_{c} = 0.038(1)$.  The cusp feature is either significantly muted or absent for the aligned counterparts.

The appearance of a central hole for $\tilde{r}_{1} = 0.5$ as cone shaped dimples 
above and below the equatorial plane merge 
suggests an explanation for the $\phi_{c}$ cusp in the case of randomly 
oriented grains with circular cross section.  The sharp increase in $\phi_{c}$ as one approaches $\tilde{r}_{1} = 0.5$ from the left
is consistent with the upper and lower concave depressions being sheltered pockets of space, which enhances the critical porosity fraction.
On the other hand, immediately to the right of the $\tilde{r} = 0.5$ boundary, the incipient central hole serves as as channel uniting 
the upper and lower voids, abruptly enhancing the overall permeability of the system and hence accounting for the sudden drop in 
$\phi_{c}$ with increasing $\tilde{r}_{1}$. 

The critical porosity for aligned tori with circular cross sections does not exhibit a prominent cusp as in the case of the randomly 
oriented counterparts.  In addition, as $\tilde{r}_{1}$ increases toward unity (i.e. in the regime of ring-like or high aspect ratio
grains), $\phi_{c}$ for aligned toroidal inclusions ultimately saturates at a markedly higher  porosity fraction than for 
randomly oriented tori.
A possible explanation for both the absence of a cusp at the topological transition and convergence to a higher $\phi_{c}$ as aligned tori 
become ring-like is the tendency of neighbors above and below to shield the cone shaped dimples or central hole for $\tilde{r}_{1} < 1/2$ and 
$\tilde{r}_{1} > 1/2$ respectively, thereby offsetting the potential enhancement to overall permeability offered by the emergence of a conduit 
through the grain.  Similar reasoning may also explain why $\phi_{c}$ for aligned grains exceeds that of randomly oriented counterparts in 
the high aspect ratio regime; again, the alignment of the axes of symmetry of the tori may increase the likelihood of the grain's central 
hole being closed off from connected void spaces, favoring a lower critical concentration than for the randomly oriented counterparts.

The inset of Fig.~\ref{fig:Fig6} shows critical porosities in the case of toroidal grains with a square 
cross section with open diamonds and filled squares representing $\phi_{c}$ results for randomly oriented and aligned inclusions, respectively.
For $r_{1}^{*}$ in the vicinity of $1/\sqrt{5}$ with an incipient cylindrical central channel, 
$\phi_{c}$ for systems made up of aligned grains exceeds that of media comprised 
of randomly oriented grains.  However, this ordering
is reversed with increasing $r_{1}^{*}$ with a crossing in the $\phi_{c}$ curves near $r_{1}^{*} = 0.75$.  Ultimately, the critical 
porosity fraction saturates beyond 0.060 if the impermeable inclusions are aligned and near 0.053(2) for randomly oriented grains.  

In contrast to the case of tori of circular cross section, there is no cusp feature for either aligned or randomly
oriented grains of square cross section.  The absence of a sharp decrease in $\phi_{c}$ with increasing $r_{1}^{*}$ 
may be due to the abrupt appearance of a central hole with no cone shaped indentations above or below the equatorial 
plane.  In the absence of these void pockets the emerging central channels thus have a significantly diminished role in expanding 
navigable connected volume networks.

\begin{figure}
\includegraphics[width=.4\textwidth]{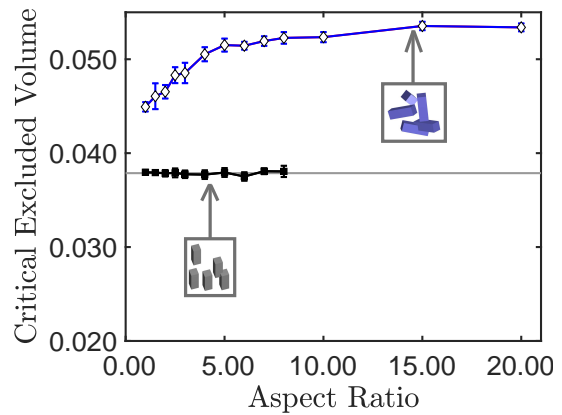}
\caption{\label{fig:Fig7} (Color online) Critical Excluded volumes for aligned (filled square symbols) and rotated (open diamonds) square prisms}
\end{figure}

\begin{figure}
\includegraphics[width=.45\textwidth]{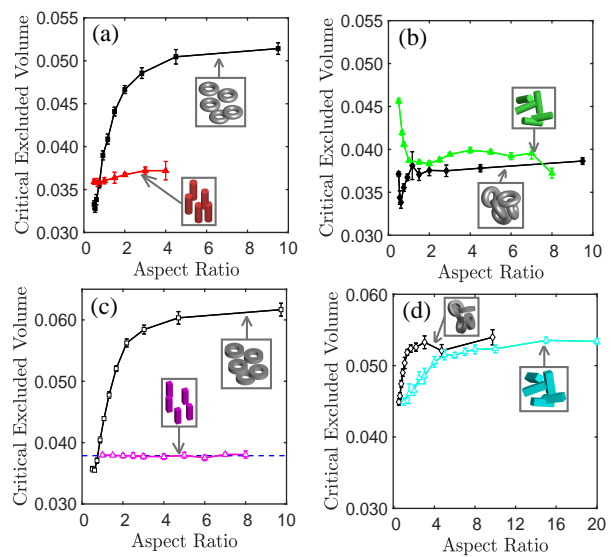}
\caption{\label{fig:Fig8} (Color online) Critical porosity fraction versus aspect ratio for aligned tori of circular cross section and 
cylinders [filled squares and triangles panel (a)], randomly oriented tori of circular cross section and cylinders
[filled diamonds and triangles in panel (b)], aligned tori of square cross section and square prisms [open squares and triangles in panel (c)],
and randomly oriented tori of square cross section and square prisms [open diamonds and triangles in panel (d)].}   
\end{figure}

As $\tilde{r}_{1}$ and $r_{1}^{*}$ approach unity with ringlike shapes for tori of circular and square cross section one also 
attains the high aspect ratio regime where small segments of toroidal grains resemble the linear counterparts, cylinders and square prisms
respectively. Although percolation thresholds for aligned and randomly oriented cylinders have been previously  
calculated~\cite{djpriour2}, to our knowledge 
critical indices not been reported for  elongated (non-cubic) grains, and are calculated in this work;
results for a variety of aspect ratios $r_{\mathrm{A}}$ appear in Tables~\ref{tab:Tab5} and~\ref{tab:Tab6}  with $\phi_{c}$ 
displayed in the graph in Fig.~\ref{fig:Fig7}.  Filled squares and open diamonds represent aligned and randomly
oriented grains respectively. While $\phi_{c}$ in the case of randomly oriented grains increases monotonically with $r_{\mathrm{A}}$, ultimately 
saturating in the vicinity of $0.052(5)$, the critical porosity fraction for aligned counterparts appears (at least up to Monte Carlo error) to 
be constant at 0.0379(5) (the broken horizontal line indicates this mean value) 
with respect to the aspect ratio (for a significant range of $r_{\mathrm{A}}$ values) 
of prism shaped inclusions. The markedly lower $\phi_{c}$ values for the aligned case may be
due to the presence of channels flanked and sheltered by parallel rectangular facets, uninterrupted void conduits which 
tend to be disrupted in the case of randomly oriented square prisms.

Among toroidal grains and linear counterparts of both circular and square cross sections, as is evident in 
panel (b) and panel (d) of Fig.~\ref{fig:Fig8}, 
the critical porosities and corresponding linear inclusions for randomly oriented converge. On the other hand, as may be 
seen in panels (a) and (c) of Fig.~\ref{fig:Fig8}, 
for aligned inclusions critical porosities are smaller for toroidal grains than for linear counterparts in the small aspect ratio regime
where the tori have a more compact structure. However, the $\phi_{c}$ curves cross for $r_{\mathrm{A}} \sim 1$ with the critical
porosities for tori rising monotonically and eventually saturating at a significantly higher value than for the linear counterparts. 
The variation of the orientation of small quasi-linear segments of ring-like tori in the equatorial plane 
likely disrupts long narrow channels which remain intact for aligned cylinders and prisms in the high aspect ratio regime, thereby 
elevating $\phi_{c}$ for aligned tori.

\subsection{Validation of Dynamical Infiltration Simulations}

\begin{figure}
\includegraphics[width=.45\textwidth]{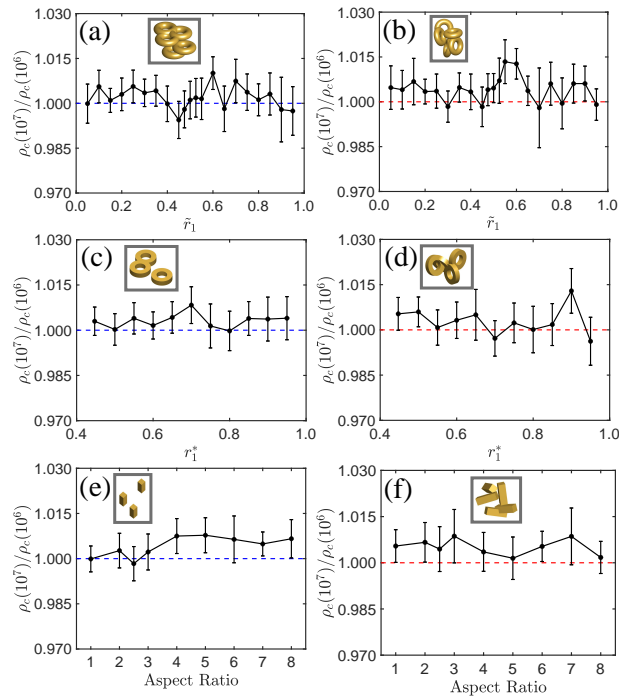}
\caption{\label{fig:Fig9} (Color online) Critical concentration ratios for calculations with $10^{7}$ and $10^{6}$ interactions of 
virtual tracer particles with grains.  Panels on the left and right correspond to aligned and randomly oriented grains respectively. 
Results for tori of circular cross section appear in panels (a) and (b), for tori of square cross section in panels (c) and (d), and 
for square prisms in panels (e) and (f).} 
\end{figure}

\begin{figure}
\includegraphics[width=.45\textwidth]{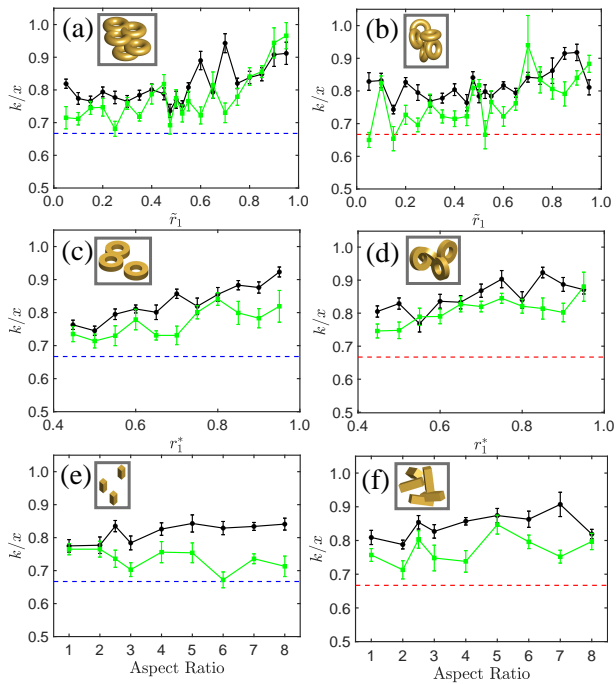}
\caption{\label{fig:Fig10} (Color online) Exponent ratios $k/x$ with filled black circles and filled green squares corresponding to 
calculations for $10^{7}$ and $10^{6}$ interactions of virtual tracer particles with grains, respectively. Panels on the left 
correspond to aligned grains, and those on the right to randomly oriented inclusions.  Results for tori of circular 
cross section are shown in panels (a) and (b), for tori of square cross section in panels (c) and (d), and for 
square prisms in panels (e) and (f). Broken horizontal lines indicate the 0.667 ratio based on universal scaling arguments.}
\end{figure}

To validate dynamical infiltration for the calculation of the grain concentration $\rho_{c}$ at the percolation threshold, 
we have conducted studies in which virtual tracers interact
with constituent impermeable grains a mean of $10^{6}$ times (i.e. ``short'' runs) with constituent impermeable grains as well as  
dwell times an order of magnitude longer (referred to here as  ``long'' runs) with 
at least on average $10^{7}$ collisions of tracers with inclusions.  The graphs in Fig.~\ref{fig:Fig9} display ratios of $\rho_{c}$ 
gleaned from long runs to those obtained from short runs for each geometry considered in this work.   
In general, the ratios are consistent with unity (indicated by broken horizontal lines) up to Monte Carlo statistical error.

Dynamical exponent ratios $k/x$ are plotted in the graphs in Fig.~\ref{fig:Fig10} for each grain geometry
considered in this work with circles and squares indicating long and short run 
results, respectively; again, leftmost panels pertain to aligned grains while panels on the right correspond to randomly oriented 
inclusions.  The broken horizontal lines indicate the $0.667$ exponent ratio obtained from universal scaling arguments.  With trajectories 
encompassing an order of magnitude more interactions with impermeable grains, $k/x$ results are generally in closer agreement with the 
universal scaling result.  Based on this trend, one may surmise that increasing the dwell  time even further would yield $k/x$ results 
in even closer accord with the 0.667 line.  Nevertheless, as may be seen in Fig.~\ref{fig:Fig9}, 
critical porosities $\phi_{c}$  are converged with respect to the dynamical infiltration trajectory length.  

\subsection{Tables of Critical Indices}

\begingroup
\begin{table}[h]
\centering
\begin{tabular}{| c | c | c | c | c | c |}
\hline
$\tilde{r}_{1}$ & $\phi_{c}$ & $k_{\mathrm{coll}}$ & $k_{\mathrm{cross}}$ & $x_{\mathrm{coll}}$ \\
\hline
\hline
0.05 & 0.0301(5) & 0.18(1) & 0.17(1) & 0.25(1) \\
\hline
0.10 & 0.0298(3) & 0.16(1) & 0.18(1) & 0.23(1) \\
\hline
0.15 & 0.0303(3) & 0.18(1) & 0.19(1) & 0.24(1) \\
\hline
0.20 & 0.0303(4) & 0.18(1) & 0.18(1) & 0.24(1) \\
\hline
0.25 & 0.0300(3) & 0.16(1) & 0.16(1) & 0.24(1) \\
\hline
0.30 & 0.0307(2) & 0.18(1) & 0.18(1) & 0.24(1) \\
\hline
0.35 & 0.0306(2) & 0.17(1) & 0.17(1) & 0.24(1) \\
\hline
0.40 & 0.0314(4) & 0.18(1) & 0.17(1) & 0.23(1) \\
\hline
0.45 & 0.0326(5) & 0.20(1) & 0.21(1) & 0.24(1) \\
\hline
0.475 & 0.0320(5) & 0.16(1) & 0.16(1) & 0.23(1) \\
\hline
0.50 & 0.0332(5) & 0.17(1) & 0.15(1) & 0.22(1) \\
\hline
0.525 & 0.328(4) & 0.17(1) & 0.18(1) & 0.23(1) \\
\hline
0.55 & 0.0338(6) & 0.18(1) & 0.19(1) & 0.23(1) \\
\hline
0.60 & 0.0355(3) & 0.16(1) & 0.15(1) & 0.22(1) \\
\hline
0.65 & 0.0390(5) & 0.19(1) & 0.18(1) & 0.23(1) \\
\hline
0.70 & 0.0409(6) & 0.16(1) & 0.18(1) & 0.23(1) \\
\hline
0.75 & 0.0441(5) & 0.18(1) & 0.20(1) & 0.23(1) \\
\hline
0.80 & 0.0466(5) & 0.18(1) & 0.16(1) & 0.22(1) \\
\hline
0.85 & 0.0486(6) & 0.18(1) & 0.19(1) & 0.21(1) \\
\hline
0.90 & 0.0505(8) & 0.21(1) & 0.19(1) & 0.23(1) \\
\hline
0.95 & 0.0514(7) & 0.20(1) & 0.18(1) & 0.20(1) \\
\hline
\hline
\end{tabular}
\caption{\label{tab:Tab1} Critical indices for tori with circular cross sections with aligned axes of symmetry.}
\end{table}
\endgroup

\begingroup
\begin{table}[h]
\centering
\begin{tabular}{| c | c | c | c | c | c |}
\hline
$\tilde{r}_{1}$ & $\phi_{c}$ & $k_{\mathrm{coll}}$ & $k_{\mathrm{cross}}$ & $x_{\mathrm{coll}}$ \\
\hline
\hline
0.05 & 0.0297(3) & 0.16(1) & 0.16(1) & 0.24(1) \\
\hline
0.10 & 0.0302(3) & 0.18(1) & 0.17(1) & 0.21(1) \\
\hline
0.15 & 0.0295(7) & 0.15(1) & 0.18(1) & 0.22(1) \\
\hline
0.20 & 0.0305(3) & 0.17(1) & 0.18(1) & 0.24(1) \\
\hline
0.25 & 0.0307(3) & 0.17(1) & 0.16(1) & 0.24(1) \\
\hline
0.30 & 0.0315(4) & 0.18(1) & 0.19(1) & 0.24(1) \\
\hline
0.35 & 0.0318(4) & 0.17(1) & 0.18(1) & 0.23(1) \\
\hline
0.40 & 0.0331(5) & 0.17(1) & 0.18(1) & 0.24(1) \\
\hline
0.45 & 0.0345(5) & 0.17(1) & 0.15(1) & 0.23(1) \\
\hline
0.475 & 0.0358(1) & 0.17(1) & 0.16(1) & 0.21(1) \\
\hline
0.50 & 0.0371(3) & 0.16(1) & 0.16(1) & 0.20(1) \\
\hline
0.525 & 0.0344(8) & 0.15(1) & 0.15(1) & 0.22(1) \\
\hline
0.55 & 0.0338(7) & 0.18(1) & 0.19(1) & 0.23(1) \\
\hline
0.60 & 0.0355(4) & 0.16(1) & 0.15(1) & 0.22(1) \\
\hline
0.65 & 0.0367(4) & 0.17(1) & 0.16(1) & 0.23(1) \\
\hline
0.70 & 0.0381(16) & 0.21(1) & 0.18(1) & 0.22(1) \\
\hline
0.75 & 0.0371(7) & 0.17(1) & 0.17(1) & 0.20(1) \\
\hline
0.80 & 0.0376(6) & 0.17(1) & 0.19(1) & 0.22(1) \\
\hline
0.85 & 0.0375(7) & 0.17(1) & 0.16(1) & 0.21(1) \\
\hline
0.90 & 0.0378(4) & 0.17(1) & 0.18(1) & 0.21(1) \\
\hline
0.95 & 0.0386(4) & 0.19(1) & 0.21(1) & 0.21(1) \\
\hline
\hline
\end{tabular}
\caption{\label{tab:Tab2} Critical indices for randomly oriented tori with circular cross sections.}
\end{table}
\endgroup

\begingroup
\begin{table}[h]
\centering
\begin{tabular}{| c | c | c | c | c | c |}
\hline
$r_{1}^{*}$ & $\phi_{c}$ & $k_{\mathrm{coll}}$ & $k_{\mathrm{cross}}$ & $x_{\mathrm{coll}}$ \\
\hline
\hline
$1/\sqrt{5}$ & 0.0357(4) & 0.17(1) & 0.18(1) & 0.24(1) \\
\hline
0.50 & 0.0355(3) & 0.18(1) & 0.18(1) & 0.26(1) \\
\hline
0.55 & 0.0371(4) & 0.18(1) & 0.18(1) & 0.25(1) \\
\hline
0.60 & 0.0404(5) & 0.20(1) & 0.20(1) & 0.25(1) \\
\hline
0.65 & 0.0439(2) & 0.18(1) & 0.17(1) & 0.25(1) \\
\hline
0.70 & 0.0478(5) & 0.18(1) & 0.20(1) & 0.24(1) \\
\hline
0.75 & 0.0521(5) & 0.18(1) & 0.18(1) & 0.23(1) \\
\hline
0.80 & 0.0563(7) & 0.20(1) & 0.20(1) & 0.23(1) \\
\hline
0.85 & 0.0584(7) & 0.18(1) & 0.20(1) & 0.23(1) \\
\hline
0.90 & 0.0603(10) & 0.18(1) & 0.20(1) & 0.23(1) \\
\hline
0.95 & 0.0617(11) & 0.18(1) & 0.19(1) & 0.22(1) \\
\hline
\hline
\end{tabular}
\caption{\label{tab:Tab3} Critical indices for tori with square cross sections and aligned axes of symmetry.}
\end{table}
\endgroup

\begingroup
\begin{table}[h]    
\centering
\begin{tabular}{| c | c | c | c | c | c |}
\hline
$r_{1}^{*}$ & $\phi_{c}$ & $k_{\mathrm{coll}}$ & $k_{\mathrm{cross}}$ & $x_{\mathrm{coll}}$ \\
\hline
\hline
$1/\sqrt{5}$ & 0.449(4) & 0.16(1) & 0.17(1) & 0.21(1) \\
\hline
0.50 & 0.0458(4) & 0.17(1) & 0.17(1) & 0.23(1) \\
\hline
0.55 & 0.0475(5) & 0.17(1) & 0.18(1) & 0.22(1) \\
\hline
0.60 & 0.0490(5) & 0.17(1) & 0.17(1) & 0.22(1) \\
\hline
0.65 & 0.0504(8) & 0.17(1) & 0.15(1) & 0.21(1) \\
\hline
0.70 & 0.0519(4) & 0.19(1) & 0.20(1) & 0.23(1) \\
\hline
0.75 & 0.0524(5) & 0.18(1) & 0.18(1) & 0.22(1) \\
\hline
0.80 & 0.0526(6) & 0.18(1) & 0.18(1) & 0.22(1) \\
\hline
0.85 & 0.0533(9) & 0.16(1) & 0.19(1) & 0.20(1) \\
\hline
0.90 & 0.0521(9) & 0.16(1) & 0.16(1) & 0.20(1) \\
\hline
0.95 & 0.0540(11) & 0.19(1) & 0.17(1) & 0.21(1) \\
\hline
\hline
\end{tabular}
\caption{\label{tab:Tab4} Critical indices for randomly oriented tori with square cross sections.}
\end{table}
\endgroup

\begingroup
\begin{table}[h]
\centering
\begin{tabular}{| c| c | c | c | c | c|}
\hline
Aspect Ratio & $\phi_{c}$  & $k_{\mathrm{coll}}$ & $k_{\mathrm{cross}}$  &  $x_{\mathrm{coll}}$ \\
\hline
\hline
1 & 0.0380(3)  & 0.19(1) & 0.17(1)  & 0.25(1)  \\
\hline
3/2 & 0.0379(2)  & 0.19(1) & 0.19(1) & 0.26(1)  \\
\hline
2 & 0.0379(5)  & 0.19(1)  & 0.19(1) &  0.25(1)  \\
\hline
5/2 & 0.0379(5) & 0.19(1) & 0.18(1)  & 0.26(1)  \\
\hline
3 & 0.0378(4)  & 0.19(1) & 0.21(1) &  0.27(1) \\
\hline
4 & 0.0377(5) & 0.19(1) & 0.19(1) & 0.25(1)  \\
\hline
5 & 0.0379(5) & 0.19(1) & 0.20(1) &  0.26(1)  \\
\hline
6  & 0.0375(5) & 0.18(1)  & 0.18(1) & 0.27(1)  \\
\hline
7 & 0.0381(3) & 0.20(1) & 0.18(1) &  0.27(1)  \\
\hline
8  &  0.0381(6) & 0.20(1) & 0.18(1) & 0.28(1)  \\
\hline
\hline
\end{tabular}
\caption{\label{tab:Tab5} Critical indices for square prisms with axes of symmetry aligned.}
\end{table}
\endgroup

\begingroup
\begin{table}[h]
\centering
\begin{tabular}{| c| c | c | c | c | c|}
\hline
Aspect Ratio & $\phi_{c}$  & $k_{\mathrm{coll}}$ & $k_{\mathrm{cross}}$  &  $x_{\mathrm{coll}}$ \\
\hline
\hline
1 & 0.0449(5)  & 0.17(1) & 0.19(1)  & 0.23(1)  \\
\hline
3/2 & 0.0461(14)  & 0.18(1) & 0.21(1) & 0.23(1)  \\
\hline
2 & 0.0465(7)  & 0.16(1)  & 0.17(1) &  0.22(1)  \\
\hline
5/2 & 0.0483(8) & 0.16(1) & 0.17(1)  & 0.22(1)  \\
\hline
3 & 0.0485(11)  & 0.18(1) & 0.19(1) &  0.22(1) \\
\hline
4 & 0.0505(7) & 0.19(1) & 0.19(1) & 0.24(1)  \\
\hline
5 & 0.0515(7) & 0.17(1) & 0.19(1) &  0.22(1)  \\
\hline
6  & 0.0519(5) & 0.17(1)  & 0.18(1) & 0.23(1)  \\
\hline
7 & 0.0519(3) & 0.17(1) & 0.18(1) &  0.23(1)  \\
\hline
8  &  0.0523(6) & 0.18(1) & 0.18(1) & 0.22(1)  \\
\hline
10 & 0.0524(5) & 0.17(1) & 0.17(1) & 0.22(1) \\
\hline
15 & 0.0535(4) & 0.19(1) & 0.19(1) & 0.23(1) \\
\hline
20 & 0.0534(4) & 0.16(1) & 0.15(1) & 0.23(1) \\
\hline
\hline
\end{tabular}
\caption{\label{tab:Tab6} Critical indices for randomly oriented square prisms.}
\end{table}
\endgroup

\section{Conclusions}

We have examined percolation phenomena for non-convex impermeable grains in the form of 
toroidal shapes of circular and square cross sections. In the case of the former, the 
topological transition signaled by the appearance of a central hole is also marked by 
a cusp in the critical porosity fraction with incipient central holes acting as channels
to expand the network of void volumes.  We find a convergence and saturation at 
a common value of $\phi_{c}$ for the 
randomly oriented tori and linear counterparts in the high aspect ratio limit where 
cross sectional length scales are dwarfed by the central hole radius.

We have achieved significant gains in computational efficiency by implementing a new approach in which voxels 
occupied by virtual tracers as well as neighboring cells are populated with grains, which 
allows rescaling the inclusions relative to voxels.  In this manner, with an optimal choice 
of the inclusion dilation factor, the computational 
burden remains essentially constant even in the high aspect ratio regime.

\section{Appendix}

Toroidal grains of circular cross section are characterized by the orientation $\hat{u}$ of the axis of symmetry, their location $\hat{x}$, and 
the major and minor axes $r_{1}$ and $r_{2}$ with the former being the radius of revolution and the latter the cross section radius.

  It is convenient to operate in terms of the position relative to the torus geometric center, $\vec{\Delta} \equiv \vec{x} - \vec{x}_{c}$ where 
$\vec{x} = \vec{x}_{0} + \hat{v} t$ in seeking the time $t$ for the virtual tracer particle to intersect with the toroidal surface. 
The vector $\vec{v}_{\mathrm{plane}} \equiv 
r_{1}[\vec{\Delta} - (\vec{\Delta} \cdot \vec{u} ) \hat{u}]/\sqrt{ \Delta^{2} - (\vec{\Delta} \cdot \hat{u})^{2} }$ is aligned with the projection of 
$\vec{\Delta}$ onto the equatorial plane of the torus and is on the circle of radius $r_{1}$, the locus 
of centers of circles of revolution.  For an intersection with the surface of the grain, one insists 
that $| \vec{\Delta} - \vec{v}_{\mathrm{plane}} | = r_{2}$.  Squaring both sides, isolating and squaring the radical expression, and 
inserting $\vec{\Delta} = \vec{\Delta}_{0} + \hat{v} t$ 
(where $\vec{\Delta}_{0} \equiv \vec{x}_{0} - \vec{x}_{c}$) yields a quartic equation in $t$:
\begin{eqnarray} 
t^{4} &+& 4 (\vec{\Delta}_{0} \cdot \hat{v} ) t^{3}  \\ \nonumber
&+& [4 (\vec{\Delta} \cdot \hat{v})^{2} + 4 \Delta_{0}^{2} - 2 (r_{1}^{2} + r_{2})^{2} + 2 r_{1}^{2} ( \hat{v} \cdot \hat{u} )^{2} ] t^{2}  \\ \nonumber
&+& [ 4 \Delta_{0}^{2} (\vec{\Delta}_{0} \cdot \hat{v} ) - 4 (\vec{\Delta}_{0} \cdot \hat{v} )(r_{1}^{2} + r_{2}^{2})\\ 
\nonumber &+& 8 (\hat{v} \cdot \hat{u} ) 
(\vec{\Delta}_{0} \cdot \hat{u} ) r_{1}^{2} ] t \\ \nonumber 
&+& [ \Delta_{0}^{2} + (r_{1}^{2} - r_{2}^{2}) ]^{2} = 0
\label{eq:Eq1000}
\end{eqnarray}
The roots may be found analytically~\cite{neumark}, and for a computationally robust approach we introduce a shift in $t$ to eliminate the linear term in
Eq.~1 and then factor the resulting depressed quartic into two readily soluble quadratic equations.   
Fortuitously, this factorization involves solving a cubic equation, which always has at least one real root.

As in the case of tori
with square cross sections, care must be taken if a virtual tracer interacts with the same grain more than once in succession.  In such a
situation, factoring out the zero root  yields a cubic equation, which also is amenable to exact solution.  Disposing of the zero root
(whose physical meaning is that the tracer particle has not left the surface)  in this
manner avoids its confusion with a scattering event due to limits on numerical precision.

When the nearest intersection point $\vec{\Delta}$ has been identified, one then obtains the surface normal, given by $\hat{n} = \vec{v}_{\perp}/r_{2}$
where the locally perpendicular direction is $\vec{v}_{\perp} = \vec{\Delta} - \vec{v}_{\mathrm{planar}}$.


\begin{acknowledgments}
We acknowledge helpful discussions with Michael Crescimanno.
Calculations in this work have benefited from use of the Ohio Supercomputer facility (OSC)~\cite{OSC}.
\end{acknowledgments}


\end{document}